\documentclass[12pt]{article}
\usepackage{amssymb,amsfonts,amsmath}
\usepackage{epsfig}
\usepackage{graphicx} 

\newcommand{\be}{\begin{equation}}
\newcommand{\ee}{\end{equation}}

\newcommand{\ba}{\begin{eqnarray}}
\newcommand{\ea}{\end{eqnarray}}

\date{}
\begin{document}

\title{\bf Exotic meson decays in the environment with chiral imbalance\footnote{Based on Talks given at XXIII International Workshop
on High Energy Physics and Quantum Field Theory, Yaroslavl', June 26 - July 3, 2017 and at IV RUSSIAN-SPANISH CONGRESS: Particle, Nuclear, Astroparticle Physics and Cosmology, JINR Dubna,
September 4-8, 2017.
}}
\vspace{1cm}
\author{A. A. Andrianov$^{1,2,}$\footnote{E-mail:andrianov@icc.ub.edu},
V. A. Andrianov$^{1,}$\footnote{E-mail:v.andriano@rambler.ru},
D. Espriu$^{2,}$\footnote{E-mail:espriu@icc.ub.edu}, \\
A. V. Iakubovich$^{1,}$\footnote{E-mail:
a.yakubovich@outlook.com},
A. E. Putilova$^{1,}$\footnote{E-mail:
alyput@gmail.com}\\
\\
$^{1}$ {\small Faculty of Physics, Saint Petersburg State
University, Universitetskaya nab. 7/9,}\\
 {\small Saint Petersburg 199034, Russia}
\\
$^{2}$ {\small Departament de F\'isica Qu\`antica i Astrof\'isica and Institut de Ci\'encies del Cosmos (ICCUB),}\\
{\small Universitat de Barcelona, Mart\'i i Franqu\'es 1, 08028 Barcelona, Spain}}

\maketitle

\abstract{ An emergence of Local Parity Breaking (LPB) in central
heavy-ion collisions (HIC) at high energies is discussed. LPB in the
fireball can be produced by a difference between the number
densities of right- and left-handed chiral fermions (Chiral
Imbalance) which is implemented by a chiral (axial) chemical
potential. The effective meson lagrangian induced by  QCD is
extended to  the medium with Chiral Imbalance and the properties of
light scalar and pseudoscalar mesons ($\pi, a_0$) are analyzed. It
is shown that exotic decays of scalar mesons arise as a result of
mixing of $ \pi $ and $ a_0 $ vacuum states  in the presence of chiral imbalance. The pion electromagnetic formfactor obtains an unusual parity-odd supplement which generates a photon polarization asymmetry in pion
polarizability. We hope that the above pointed indications of LPB
can be identified in experiments on LHC, RHIC, CBM FAIR and NICA
accelerators.}

\maketitle

\section{ Topological charge, Chiral Imbalance and axial chemical potential}

The behaviour of baryonic matter under
extreme conditions has got recently a lot of interest \cite{1,2}. A medium
generated in the heavy ion collisions may serve for detailed  studies, both
experimental and theoretical, of various phases of hadron
matter.  In this context new properties of QCD in the hot and dense environment
are tested in current accelerator experiments on RHIC and LHC
\cite{3, 4}.

In heavy ion collisions, in principle, there are two distinct
experimental situations for peripheral and central collisions.
In the first case  the so-called Chiral Magnetic Effect (CME) can be detected, details see in
~\cite{kpt} and also ~\cite{8} for a review and additional references.

In the second case there are some experimental indications of an abnormal
dilepton excess in the range of low invariant masses and rapidities
and moderate values of the transverse momenta
~\cite{ceres}--\cite{lapidus} (see the reviews in ~\cite{tser}),
which can be thought of as a result of LPB
in the medium (the details can be found in ~\cite{tmf}). In
particular, in heavy-ion collisions at high energies, with raising
temperatures and baryon densities, metastable states can appear
in the finite-volume fireball with a nontrivial topological axial charge
(due to fluctuations of gluonic fields) ~$T_{5}$, which is
related to the gluon gauge field ~$G_{i}$,
\begin{equation}
T_5(t)=\frac{1}{8\pi^2}\int_{\mbox{\rm  vol.}}d^3x\,\varepsilon_{jkl}\,
\mbox{\rm  Tr}\biggl(G^j\partial^k
G^l-i\frac{2}{3}G^jG^kG^l\biggr),\quad j,k,l = 1,2,3, \label{eq20}
\end{equation}
where the integration is over the fireball volume.
Its jump ~$\Delta T_{5}$ can be
associated with the space-time integral of the gauge-invariant
Chern-Pontryagin density,
\begin{align}
\aligned \Delta T_5&=T_5(t_f)-T_5(0)=\frac{1}{16\pi^2}\int^{t_f}_0
dt\int_{\text{vol.}}d^3x\,\mbox{\rm  Tr} (G^{\mu\nu}\widetilde
G_{\mu\nu})=
\frac{1}{4\pi^2}\int^{t_f}_0 dt\int_{\text{vol.}}
d^3x\,\partial^\mu K_\mu,
\\
K_\mu&=\frac{1}{2} \epsilon_{\mu\nu\rho\sigma}\mbox{\rm  Tr}
\biggl(G^\nu\partial^\rho G^\sigma- i\frac{2}{3}G^\nu G^\rho
G^\sigma\biggr).
\endaligned
\label{eq21}
\end{align}
For the time being we adopt a static case and neglect a topological current flux through the fireball boundary
during lifetime of the corresponding thermodynamic phase in a
domain.

It is known that the divergence of isosinglet axial quark current
$J_{5,\mu}=\overline q \gamma_\mu\gamma_5 q$ is locally constrained
via the relation of partial conservation of axial current  affected
by the gluon anomaly,
\begin{equation}
\partial^\mu J_{5,\mu}-2i \widehat m_q J_5 =
\frac{N_f}{2\pi^2}\partial^\mu K_\mu;\quad J_5 = \overline q \gamma_5 q \label{eq22}
\end{equation}
This relation allows to find the relation of a nonzero topological
charge with a non-trivial quark axial charge ~$Q^q_{5}$.
Namely, integrating
over a finite volume of fireball we come to the equality,
\begin{equation}
\label{eq23}
\begin{aligned}
&\frac{d}{dt}(Q_5^q-2 N_f T_5) \simeq 2i\int_{\text{vol.}}d^3x\,
\widehat m_q \overline q\gamma_5q,
\\
&Q_5^q=\int_{\text{vol.}}d^3x\,q^\dagger\gamma_5 q = \langle
N_{L}-N_{R}\rangle,
\end{aligned}
\end{equation}
where $\langle N_{L}-N_{R}\rangle$ stands for the vacuum averaged
difference between left and right chiral densities of baryon number.
Therefrom  it follows that in the chiral limit (when the masses of
light quarks are taken zero)  the
axial quark charge is conserved in the presence of non-zero
(metastable) topological charge. If for the lifetime  of fireball
and the size of hadron fireball of order $L=5-10$~fm , the average
topological charge is non-zero, $\langle \Delta T_5 \rangle\ne 0$,
then it may be associated with a topological chemical potential
$\mu_T$
or an axial chemical potential~$\mu_5$ ~\cite{aaep} for neglected
masses of light~$u$,~$d$ quarks. Thus we have,
\begin{equation}
\langle \Delta T_5 \rangle \simeq \frac{1}{2N_f} \langle
Q_5^q\rangle\, \Longleftrightarrow\,\mu_5 \simeq
\frac{1}{2N_f}\mu_T, \label{eq24}
\end{equation}
Thus adding to the QCD lagrangian the term $\Delta{\mathcal
L}_{\mathrm{top}}=\mu_T\Delta T_5$ or $\Delta{\mathcal L}_q=\mu_5
Q_5^q$, we get the possibility of accounting for non-trivial
fluctuations of topological charge (fluctons) in the nuclear (quark) fireball.

In the general, Lorentz covariant form the field dual to the fluctons is
described by means of the classical pseudoscalar field~$a(x)$, so
that,
\begin{equation}
\Delta {\mathcal L}_{a}=\frac{N_f}{2\pi^2} K_\nu \partial^\nu a(x)\simeq \frac{1}{\pi^2} K_\nu b^\nu
\,\Longleftrightarrow\, b^\nu
\overline q \gamma_\nu\gamma_5 q,\quad b_\nu \simeq
\langle\partial_\nu a(x)\rangle \simeq{\rm const}. \label{eq25}
\end{equation}
Thus in a quasi-equilibrium situation the appearance of a nearly
conserved chiral charge can be incorporated with the help of an axial (chiral) vector  chemical potential $b_\nu$. The appearance of a space vector part in $b_\nu$ can be associated with the non-equilibrium axial charge flow \cite{kharzeev, teryaev, land}.

\section{QCD-inspired effective meson Lagrangian for $ SU_f (2)$ case}

For the detection of Local Parity Breaking in the hadron fireball
we implement the generalized sigma model with a background 4-vector of axial chemical potential \cite{sigma}, symmetric under
$ SU_{L}(N_{f})\times SU_{R}(N_{f})$, for $ u,d$-quarks ($N_{f}=2$),
\begin{align}
L&=\cfrac{1}{4}\,\mbox{\rm Tr}\,(D_{\mu}H\,(D^{\mu}H)^{\dagger})
+\cfrac{B}{2}\,\mbox{\rm Tr}\,[ \,m(H\,+\,H^{\dagger})]
+\cfrac{M^{2}}{2}\,\mbox{\rm Tr}\,(HH^{\dagger})
\nonumber \\
\,&-\cfrac{\lambda_{1}}{2}\,\mbox{\rm Tr}\,\left[(HH^{\dagger})^{2}\right]
-\cfrac{\lambda_{2}}{4}\,[\,\mbox{\rm Tr}\,(HH^{\dagger})]^{2}
+\cfrac{c}{2}\,(\det H+\,\det H^{\dagger}),
\label{lagr_sigma}
\end{align}
where $H = \xi\,\Sigma\,\xi$ is an operator for meson fields,
\(m\) is an average mass of current quarks,
\(M\) is a "tachyonic" mass generating the spontaneous breaking of chiral symmetry,
\(B, c,\lambda_{1},\lambda_{2}\) are real constants.

The matrix \(\Sigma\) includes the singlet scalar meson \(\sigma\), its vacuum average \(v\) and the isotriplet of scalar mesons \( a^{0}_{0},a^{-}_{0},a^{+}_{0}\),
\begin{gather}
\Sigma = \begin{bmatrix}v+\sigma+a^{0}_{0}\ & \sqrt{2 \,}\,a^{+}_{0} \\
\sqrt{2 \,}\,a^{-}_{0} & v+\sigma -a^{0}_{0} \\
\end{bmatrix}.
\end{gather}
The operator \(\mathbf \xi\) realizes a nonlinear representation of the chiral group and is determined by the isotriplet \( \pi^{0},\pi^{-},\pi^{+}\)
of pseudoscalar mesons,
\begin{gather}
\xi = \exp\left(\!\cfrac{i\,\vec\pi \vec\tau\ }{2\,f_{\pi}}\right)
\approx 1 + \cfrac{i\,\vec \pi\vec \tau}{2\,f_{\pi}}
\,-\,\cfrac{(\vec\pi \vec\tau)^2}{8\,f^{\,2}_{\pi}},
\\
\vec\pi \vec\tau = \begin{bmatrix}\pi^{0}\ & \sqrt{2 \,}\,\pi^{+} \\
\sqrt{2 \,}\,\pi^{-} &  -\pi^{0} \\
\end{bmatrix},
\label{pi_tau}
\end{gather}
where $\vec\tau$ are Pauli matrix,
\(f_{\pi}\) is a decay constant of $\pi$ mesons.

The covariant derivative of $H$ contains external gauge fields
${R}_{\mu}$ and ${L}_{\mu}$,
\begin{equation}
D_{\mu}H=
\partial_{\mu}H-i{L}_{\mu}H+iH{R}_{\mu}
\label{D_mu}
\end{equation}
These fields include the photon field \(A_{\mu}\) and are supplemented also a background 4-vector of axial chemical potential \( (b_\mu) = (b_0, {\mathbf b})\),
\begin{gather}
{R}_{\mu} = e\,Q_{em}A_{\mu}- b_\mu \cdot 1_{2\times2},
\nonumber \\
{L}_{\mu} = e\,Q_{em}A_{\mu}+b_\mu \cdot 1_{2\times2},
\label{ext_L_R}
\end{gather}
where
\(Q_{em} = \frac{1}{2}\tau_{3}+\frac{1}{6}1_{2\times2}\)
is a matrix of electromagnetic charge.

The complete effective meson lagrangian has to include a P-odd part the Wess-Zumino-Witten effective action \cite{WZW} which is modified in the chirally imbalanced medium. The relevant parts of WZW action read,
\begin{equation}
\Delta{\cal  L}_{WZW}= -\frac{ie\,N_{c} b_\nu }{6\pi^{2}\,v^{2}}\,\epsilon^{\;\nu\sigma\lambda\rho}\,
A_{\rho}(\partial_{\sigma}\pi^{+})\,(\partial_{\lambda}\pi^{-})
-\frac{e^{2}N_{c}}{24\,\pi^{2}v}\,\epsilon^{\,\nu\sigma\lambda\rho}\,
(\partial_{\sigma}A_{\lambda})(\partial_{\nu}A_{\rho})\pi^{0}
\end{equation}

\section{Chiral (scalar) condensate depending on chiral chemical vector}
The mass gap equation for the scalar condensate follows from (\ref{lagr_sigma}),
\begin{equation}
-4\left(\lambda_{1}+\lambda_{2}\right)v^3
+\left(2M^2+4 b^2+2c\right)v+2B\,m=0 .\nonumber
\label{massgap}
\end{equation}
The general solution of this equation reads
\begin{eqnarray}
v(b_\mu) &=&  \frac{1}{6^\frac{2}{3}(\lambda_{1}+\lambda_{2})}\,
 \Bigg[ 9B\:m(\lambda_{1}+\lambda_{2})^{2} \nonumber \\
 &+& \sqrt{3(\lambda_{1}+\lambda_{2})^{3}\left(27B^2m^2(\lambda_{1}+\lambda_{2})-2
\left(M^2+2b^2+c\right)^3\right)} \Bigg]^{1/3} \nonumber \\
 &+&   \,\frac{M^2+2b^2+c}{6^{\,^{1}_{3}}}\,
 \Bigg[ 9B\,m(\lambda_{1}+\lambda_{2})^{2} \nonumber \\
&+& \sqrt{3(\lambda_{1}+\lambda_{2})^{3}\left(27B^2m^2(\lambda_{1}+\lambda_{2})-2
\left(M^2+2b^2+c\right)^3\right)} \Bigg]^{-1/3}. \nonumber \label{exsolution}
\end{eqnarray}

There are different regions for chiral vector covariant under Lorentz transformations of fireball frame.
\begin{enumerate}
 \item Chiral charge imbalance region when $b^2 > 0$, in the rest frame the chiral background $(b^\mu) = (\mu_5, 0,0,0)$.
\item Chiral vector imbalance region with $b^2 < 0$, in the static frame the chiral background is taken along the beam axis $(b^\mu) = (0,0,0,b)$.
\item Transition region with $b^2 = 0$, in the light-cone background
$(b^\mu) = (b,0,0,\pm b)$.
\end{enumerate}

 We stress that in a hot medium Lorentz invariance is broken by thermal bath and the physical effects depend on a particular set of components of $(b^\mu)$.

The plots for condensates
display the enhancement of CSB and the restoration of chiral symmetry depending on the sign of $b_{\mu}b^{\mu}$.
\begin{figure}[h]
\vspace{-6cm}
\begin{minipage}[h]{0.49\linewidth}
\centering{\includegraphics[width=0.87\linewidth]{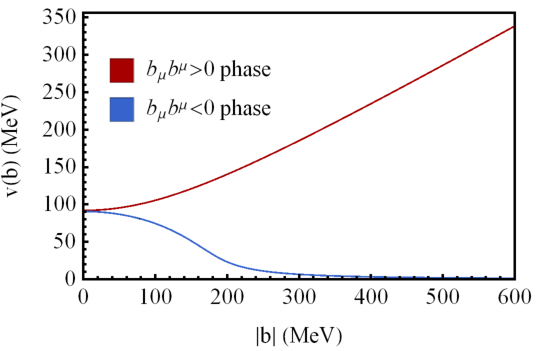}}
\vspace{-4.0cm}
\end{minipage}
\hfill
\begin{minipage}[h]{0.49\linewidth}
\centering{\includegraphics[width=1.1\linewidth]{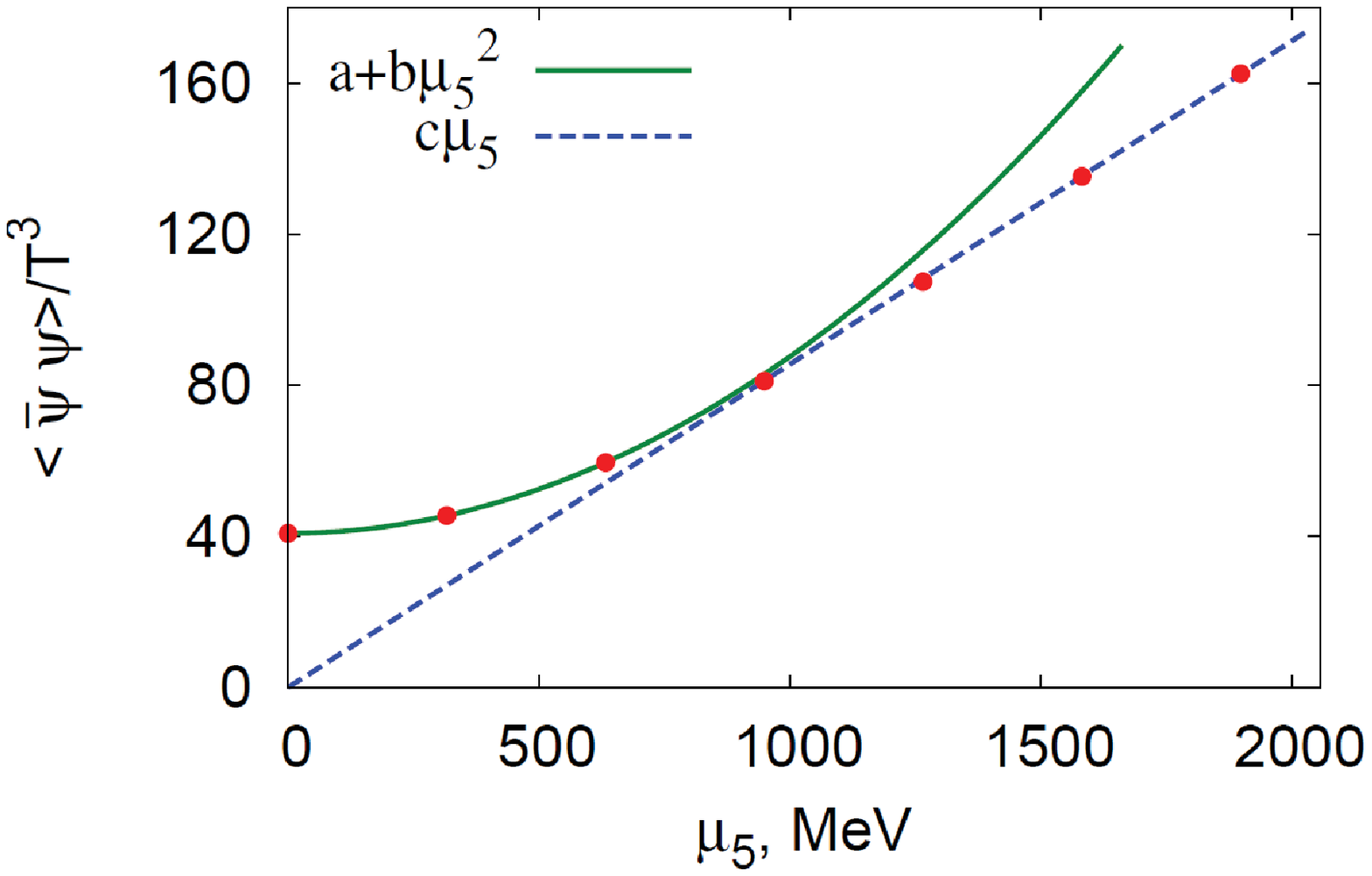}}
\vspace{+1.8cm}
\end{minipage}
\vspace{-2.5cm}
\caption{Left: the condensate in $b_{\mu}b^{\mu}>0$ and $b_{\mu}b^{\mu}<0$ region;\, right: the vacuum average $v(\mu_{5})\sim\langle\bar \psi\psi\rangle/B$ from the lattice calculations \cite{braguta}.}
\label{fig1}
\end{figure}

 Namely, in the chiral imbalance region with $b^2 > 0$ the increasing of chemical potentials causes the growth of chiral condensate, i.e. enhances Chiral Symmetry Breaking (CSB).\\
  Instead, in the chiral vector imbalance region with $b^2 < 0$ the chiral condensate is decreasing with growing $|b^2|$ up to
$|b^2| =\frac12 (M^2+c) $. At this scale in the chiral limit $m\to 0$ the CSB parameter $v \to 0$ (see below) and spontaneous CSB is restored.\\
At last the CSB parameter $v$ is insensitive to the light-cone $b^\mu$.

For $b^2 > 0$ in the rest frame of vector $(b_{\mu})= (\mu_5,0,0,0)$ one can compare  the predictions of our effective meson lagrangian (see Fig.1, the left plot) with the lattice estimations \cite{braguta} (see Fig.1, the right plot) which clearly shows the enhancement of CSB.

However non-zero real space components of $(b_{\mu})$ produce non-Hermitian purely imaginary vertices after euclidization of QCD which makes it difficult to compute on the lattice their contribution to the quark determinant.

\section{Meson mass spectrum in different chiral imbalance regions}

Introduce the definitions for meson state masses in the chiral imbalance environment.
The mass matrix for scalar and pseudoscalar mesons on the diagonal takes the following values, 
\begin{align}
m^{2}_{a}&=-2\left(M^2-2\left(3\lambda_{1}+
\lambda_{2}\right)v^2-c+2 b^{2}\right),
\nonumber\\
m^{2}_{\sigma}&=-2\left(M^2-6\left(\lambda_{1}
+\lambda_{2}\right)v^2+c+2b^{2}\right),
\nonumber\\
m^{2}_{\pi}&=\frac{2B\,m}{v}.
\end{align}
After diagonalization we define distorted masses as  $ m_{eff+}$ for the field $ \tilde a$ and $ m_{eff-}$ for the field $ \tilde \pi $
\begin{gather}
m^{2}_{eff\pm}=\frac{1}{2}\left( m^{2}_{a}+m^{2}_{\pi} \, \pm \, \sqrt{\left( m^{2}_{a}-m^{2}_{\pi} \,  \right)^{2}+(8\,b^\mu\,k_\mu)^{2}}  \right). \label{effmassgen}
\end{gather}

\subsection{Parameters of QCD-inspired generalized sigma model }

Let us normalize the vacuum parameters of our model.
We take $ m_{\pi}=139 $ MeV, $ m_a=980 $ MeV, $ m_\sigma=500 $ MeV, $ m=5.5 $ MeV, $ \mu_5=0 $, $ M=300 $ MeV, $ v=92 $  MeV. Then from the following Eqs.
\begin{equation}
\left\{
\begin{aligned}
m^{2}_{\sigma} &=-2\left( M^{2}-6\,(\lambda_{1}+\lambda_{2})v^{2}+c \,\right)\\
m^{2}_a &=-2 \left( M^{2}-2\,(3\lambda_{1}+\lambda_{2})v^{2}-c \, \right)\\
m^{2}_{\pi} &=\frac{2\,B\,m}{v}\\
v(\mu_{5}=0) &=\sqrt{\dfrac{M^{\,2}+c}{2(\lambda_{1}+\lambda_{2})}}
+\frac{B\,m}{2(M^{2}+c)}
\end{aligned}
\;\ ,
\right.
\end{equation}
 one can find $\lambda_{1}$, $\lambda_{2}$, $c$ and $b$. We have for parameters
 $\lambda_{1}=1.64850\times10 $, $ \lambda_{2}=-1.31313\times10 $,
$ c=-4.46874\times10^4 $  $ \mbox{\rm MeV}^2 $, $ B  =1.61594\times10^5 $ $\mbox{\rm MeV}^2 $ .
\subsection{Diagonalization matrix}

After diagonalization of mass matrix the states for $ a_0 $ and $\pi$ mesons happen to be mixed.
The eigenstates are defined as
\begin{gather}
a_0= C_{a \tilde a} {\tilde a} + C_{a\tilde\pi} \tilde\pi,\quad \pi= C_{\pi \tilde a}\tilde{a} + C_{\pi\tilde\pi}\tilde\pi , \nonumber \\ C_{a\tilde a}=iC_{\pi\tilde\pi}=C_+, \quad C_{a\tilde\pi}=-iC_{\pi\tilde a}=-C_-,
\end{gather}
with
\be
C_\pm=\frac 1{\sqrt 2}\sqrt{1\pm\frac{m_a^2-m_\pi^2}{\sqrt{(m_a^2-m_\pi^2)^2+(8 \,b^\mu\,k_\mu )^2}}}.
\ee
We use the notation $ \tilde a,\tilde\pi$, indicating that these states tend to $a_0$, $\pi$ when
$b^\mu=0$.

One can see (Fig.2, left plot) that for \(b_{\mu}b^{\mu}>0\) the growing chemical potential very quickly enforces the mixing between pions and isotriplet scalars so that distorted scalars happen to be involved into typical reactions of pion decay and pion formfactor. Yet with increasing of chemical potential the content of original pion states in distorted scalars and pseudoscalars diminishes.

\begin{figure}[!h]
\begin{minipage}[h]{0.49\linewidth}
\center{\includegraphics[width=0.87\linewidth]{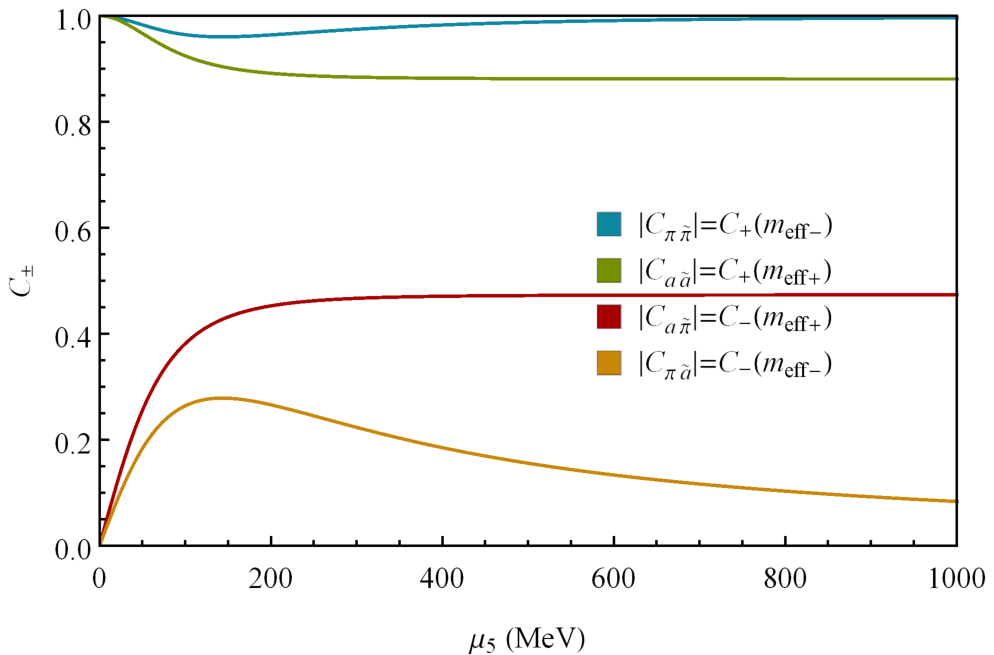}}
\end{minipage}
\hfill
\begin{minipage}[h]{0.49\linewidth}
\center{\includegraphics[width=0.9\linewidth]{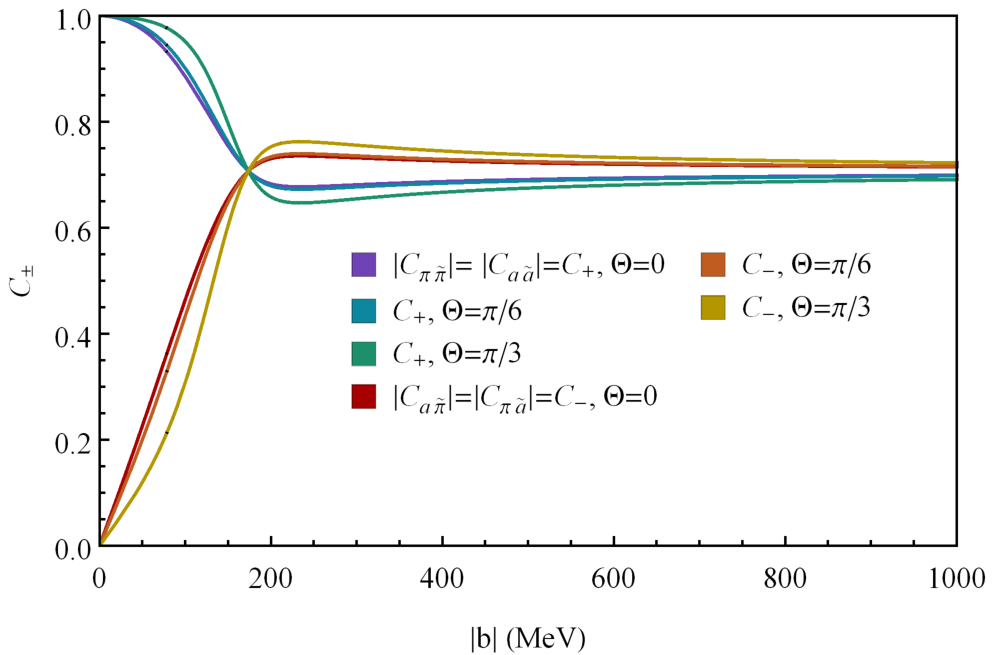}}
\end{minipage}
\caption{Left: mixing coefficients dependence on chemical potential \(\mu_{5}\) for \(|\vec k|=1000\) MeV in \(b_{\mu}b^{\mu}>0\) region;\, right: mixing coefficients dependence on chemical potential \(|b|\) for \(|\vec k|=1000\) MeV in \(b_{\mu}b^{\mu}<0\) region .}
\end{figure}
In turn, for \(b_{\mu}b^{\mu}<0\) (Fig.2, right plot) the restoration of CS does not suppress the mixing of parity counterpartners  and for increasing chemical  potential tends to a nearly constant mixing coefficients with values in between 0 and 1.

\subsection{ Masses in CSB region with \(b_{\mu}b^{\mu}>0\)}
This is a region where CSB is enhancing  and the \(\tilde a_{0}\) and \(\sigma\) mesons  become more heavy with growing chemical potential.
\begin{figure}[!h]
\center{\includegraphics[width=0.5\linewidth]{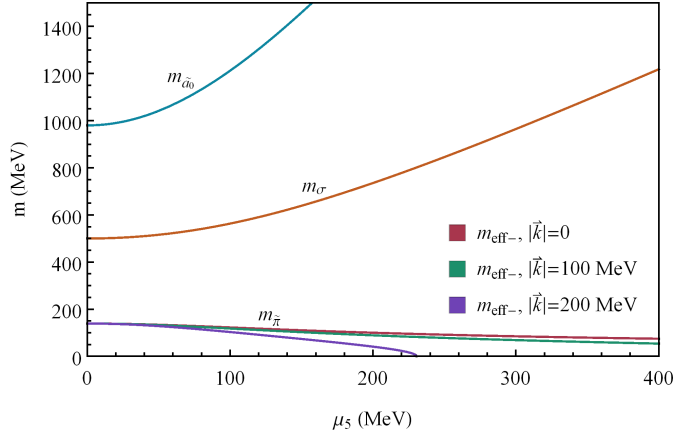}}
\caption{\(\tilde a_{0}\)-meson and \(\tilde\pi\)-meson effective mass dependence as well as  \(\sigma\)-meson mass dependence on chemical potential \( b = \mu_{5}\),  for different values of  \(|\vec k|\) in \(b_{\mu}b^{\mu}>0\) region.}
\label{}\vspace{-0.5cm}
\end{figure}

Meantime the \(\tilde\pi\)-meson effective mass is slowly decreasing at rest and decreasing faster in flight with $|\vec k|\not=0$. One can see how the \(\tilde\pi\)-meson reaches the massless point and further on its mass squared becomes negative, "tachyonic" which however does not cause any causality problems. One can check that the group velocity of these states remains less than the light velocity.
\medskip

\subsection{ Masses in CSR region with \(b_{\mu}b^{\mu}<0\)}
In this region in the rest frame one can see clearly the CS restoration with merging masses of all scalars and pseudoscalars.
\begin{figure}[!h]
\center{\includegraphics[width=0.5\linewidth]{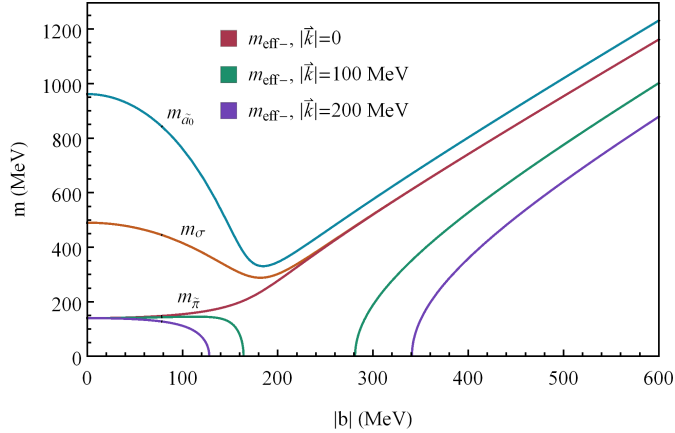}}
\caption{\(\tilde a_{0}\)-meson and \(\sigma\)-meson mass dependence on chemical potential \(b\), \(\tilde\pi\)-meson effective mass dependence on \(b\) for different values of  \(|\vec k|\) in \(b_{\mu}b^{\mu}<0\) phase, the angle between \(\vec k\) and \(\vec b\) is  \(\theta=0\)}
\label{}
\end{figure}
\medskip

But in flight the behavior of pion masses is more peculiar. First pion masses vanish and then mass squared become negative. Next they reappear with positive mass squared and slowly approach to scalar masses in asymptotics. Thus the in-flight effect of chiral vector imbalance on pion spectrum
deviates strongly from naive expectations.
\section{The decays ${\tilde a}^{\pm}_0\rightarrow \tilde\pi^{\pm}\gamma,\ \tilde\pi^0 \rightarrow \gamma\gamma,\  {\tilde a}^0_0 \rightarrow \gamma\gamma$}
After mixing $\pi, a^0_0 \rightarrow \tilde\pi, \tilde a^0_0 $ the decay ${\tilde a}^{\pm}_0\rightarrow \tilde\pi^{\pm}\gamma$ arises which breaks space parity and therefore is forbidden in vacuum.
\begin{figure}[!h]
\vspace{-0.2cm}
\centering
\includegraphics[scale=1.1]{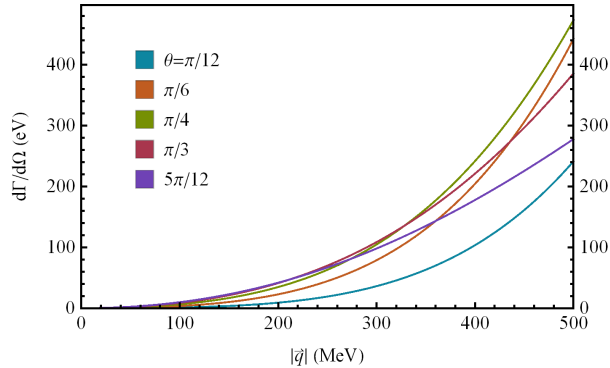}
\vspace{-0.2cm}
\caption{Decay width $a^{\pm}\rightarrow\pi^{\pm}\gamma$, $\mu_{5}=100$ MeV}
\label{fig:width1_2d}
\end{figure}

with $q =|\vec q|$ being space momenta of scalars.

As well in the decay process $\pi^{0}\rightarrow\gamma\gamma$   an adjacent resonance decay \({\tilde a}_{0}^{0}\rightarrow\gamma\gamma\) emerges after mixing. From the plots for mixing coefficients we conclude that these processes are comparable in decay widths.

\begin{figure}[!h]
\begin{minipage}[h]{0.49\linewidth}
\center{\includegraphics[width=0.9\linewidth]{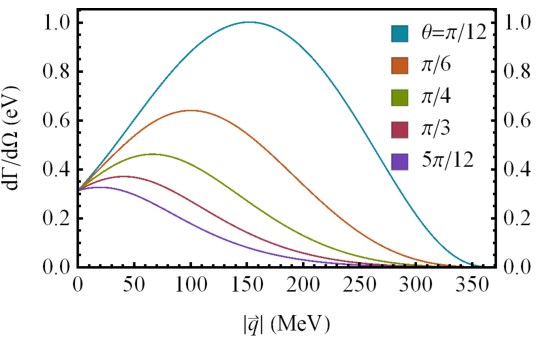}
\vspace{-0.2cm} }
\end{minipage}
\hfill
\begin{minipage}[h]{0.49\linewidth}
\center{\includegraphics[width=0.87\linewidth]{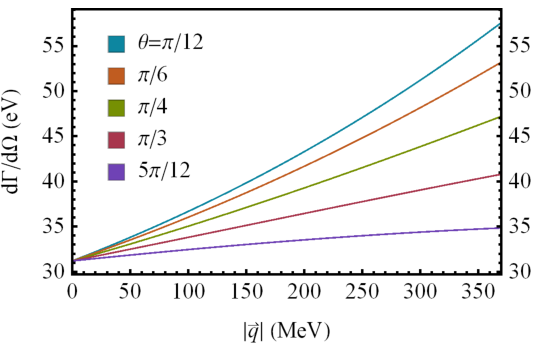} }
\vspace{-0.1cm}
\end{minipage}
\caption{Decay widths; left: $\tilde\pi^{0}\rightarrow\gamma\gamma$, right: \({\tilde a}_{0}^{0}\rightarrow\gamma\gamma\), \(\mu_{5}=100\) MeV}
\vspace{-0.5cm}
\end{figure}

For pseudoscalars and scalars in flight  the speeds of decays are considerably increasing. The effects are opposite to the Lorentz retardation.

\section{Conclusions and outlook}

In this work we described a possibility of local parity
breaking (LPB) emerging in a dense hot baryon matter (hadron fireball) in
heavy-ion collisions at high energies.
The phenomenology of LPB in a fireball is based on introducing a
topological (axial) charge and a topological (chiral) chemical
potential. Topological charge fluctuations transmit their influence to hadronic physics via an axial chemical potential. We suggested QCD-motivated sigma model for the description of isotriplet pseudoscalar and isoscalar and isotriplet scalar mesons  in the body of a fireball. We conclude:
\begin{itemize}
\item Strong CP violation is quite a challenging possibility to be revealed
in heavy-ion collisions both at high energy densities (temperatures) and being triggered by large baryon densities.
\item However the existing theoretical arguments for arising CP violation
in FINITE volumes are not well sufficient to calculate the production rate
of CP violating nuclear processes.
\item There are two ways to improve the discovery potential: firstly,
to elaborate the recipes for experimentalists to detect  peculiar effects
generated in CP-odd background, secondly, to measure the production
of  the mass states without a firm CP parity. In both cases
the chiral chemical potential method helps a lot in predictions.
\item In addition we already suggested \cite{aaep} the vector meson dominance model
with chiral imbalance:
the spectrum of massive vector mesons splits
into three components with different polarizations and with different
effective masses  that can be used to detect local parity breaking.
The proposed schemes for revealing local parity breaking helps to (partially) explain
qualitatively and quantitatively the anomalous yield of
dilepton pairs in the CERES, PHENIX, STAR, NA60, and ALICE
experiments. Accordingly the identification of its physical origin might
serve as a base for a deeper understanding of QCD properties in a
medium under extreme conditions. Experimental collaborations should
definitely check this possibility.
\item Recently an interesting proposal was given in \cite{Harada} to detect the LPB by measuring photon polarization asymmetry in the process $\pi^\pm \gamma \to \pi^\pm \gamma $. We extend this proposal  indicating the resonance enhancement at  energies comparable with the mass of $\tilde a^\pm_0$ scalars.
\end{itemize}
\section*{Acknowledgements} A.A. and V.A.
were supported by RFBR project 16-02-00348 and also got a financial support of SPbSU, in particular, by projects 11.41.415.2017, 11.41.416.2017, 11.41.417.2017, 11.42.697.2017, 11.42.698.2017, 11.41.1038.2017.
 This work has been supported through grants FPA2013-46570, 2014-SGR-104 and Consolider CPAN. Funding was also partially provided by the Spanish MINECO under project MDM-2014-0369 of ICCUB (Unidad de Excelencia `Maria de Maeztu').

\end{document}